\PassOptionsToPackage{pdfpagelabels=false}{hyperref}
\documentclass[fleqn,usenatbib,usedcolumn,letters]{mnras}
\usepackage[british]{babel}             % British English hyphenation

\usepackage{newtxmath,newtxtext}
\DeclareSymbolFont{operators}{OT1}{ntxtlf}{m}{n}
\SetSymbolFont{operators}{bold}{OT1}{ntxtlf}{b}{n}
\usepackage[T1]{fontenc}

\usepackage{graphicx}	% Including figure files
\usepackage{amsmath}	% Advanced maths commands
\usepackage{amssymb}	% Extra maths symbols
\usepackage{xspace}
\newcommand{\teff}{\ensuremath{T_\textrm{eff}}\xspace}
\newcommand{\logg}{\ensuremath{\log g}\xspace}
\newcommand{\kms}{\ensuremath{\textrm{km}\,\textrm{s}^{-1}}\xspace}
\newcommand{\elmfull}{2MASS J05005185$-$0930549\xspace}
\newcommand{\elm}{J0500$-$0930\xspace}
\newcommand{\gaia}{\textit{Gaia}\xspace}
\newcommand{\tess}{\textit{TESS}\xspace}
\newcommand{\msun}{\ensuremath{\mathrm{M}_\mathrm{\sun}}\xspace}
\newcommand{\rsun}{\ensuremath{\mathrm{R}_\mathrm{\sun}}\xspace}

\title[Closest ELM WD]{The closest extremely low-mass white dwarf to the Sun}

\author[A. Kawka et al.]{Adela~Kawka$^{1}$\thanks{Email: \texttt{adela.kawka@curtin.edu.au}},
 Jeffrey~D.~Simpson$^{2}$, St\'ephane~Vennes$^{3}$,
Michael~S.~Bessell$^{4}$, 
\newauthor
Gary~S.~Da~Costa$^{4}$, Anna~F.~Marino$^{5,6}$,
and Simon~J.~Murphy$^{7}$\\
\\
$^{1}$ International Centre for Radio Astronomy Research - Curtin University, GPO Box U1987, Perth, WA 6845, Australia\\
$^{2}$ School of Physics, UNSW Sydney, Sydney NSW 2052, Australia\\
$^{3}$ Mathematical Sciences Institute, The Australian National University,
Canberra, ACT 0200, Australia\\
$^{4}$ Research School of Astronomy \& Astrophysics, The Australian National University, Canberra, ACT 2611, Australia\\
$^{5}$ Dipartmento di Fisica e Astronomica Galileo Galilei, Universit\`a di Padova, Vicolo dell'Osservatorio 3, Padova I-35122, Italy\\
$^{6}$ Centro di Ateneo di Studi e Attivita Spaziali "Giuseppe Colombo" - CISAS, Via Venezia 15, Padova IT-35131, Italy\\
$^{7}$ School of Science, The University of New South Wales, Canberra, Canberra, ACT 2600, Australia\\
}

\date{Accepted 2020 April 11. Received 2020 April 9; in original form 2020 March 12}

\pubyear{2020}

\begin{document}
\label{firstpage}
\pagerange{\pageref{firstpage}--\pageref{lastpage}}
\maketitle

\begin{abstract}
We present the orbit and properties of 2MASS J050051.85--093054.9, establishing it as the closest ($d\approx71$~pc) extremely low mass white dwarf to the Sun. We find that this star is hydrogen-rich with $\teff\approx 10\,500$~K, $\logg\approx5.9$, and, following evolutionary models, has a mass of $\approx 0.17$~\msun. Independent analysis of radial velocity and \tess photometric time series reveals an orbital period of $\approx$9.5~h. Its high velocity amplitude ($K\approx 144~\kms$) produces a measurable Doppler beaming effect in the \tess light curve with an amplitude of 1~mmag. The unseen companion is most likely a faint white dwarf. \elm belongs to a class of post-common envelope systems that will most likely merge through unstable mass transfer and in specific circumstances lead to Type Ia supernova explosions.
\end{abstract}


\section{Introduction}
With most stars ending their lives as white dwarfs \citep[e.g.,][]{Fontaine2001}, and most stars being in binaries \citep{Moe2017}, there should be a large population of double-degenerate systems in our Galaxy. One interesting subset of these systems are extremely low mass (ELM) white dwarf (WD) binaries, where one of the stars has a mass $\leq$0.3~\msun. Such a low mass cannot be achieved through canonical single-star evolution \citep[e.g.,][]{Marsh1995}, so these ELM WDs must undergo significant mass loss during their evolution. Observationally, they are relatively rare objects \citep[$\sim$100 are known;][]{Brown2020},
and with only a few exceptions \citep{Brown2016}, they have been found to be in short-period (hours to minutes) binaries.

Beyond their interest for stellar evolution, these compact binary systems are predicted to emit gravitational waves. The magnitude of the strain measured by detectors like the future \textit{LISA} instrument \citep{AmaroSeoane2012} depend on the inverse of the distance (as well as the inclination of the system) so that the identification and characterization of the nearest systems is important.

In this work we present a detailed analysis of \elmfull (hereafter \elm) which confirms a suggestion offered by \citet{Scholz2018}, later seconded by \citet{Pelisoli2019}, that \elm is an ELM WD with an effective temperature $T_{\rm eff}\approx11,900\pm1100~{\rm K}$ and a surface gravity $\logg\leq6.5$.The measured \gaia DR2 parallax of $\varpi=13.97\pm0.05$~mas places it at a distance of $71.41\pm0.27$~pc \citep{BailerJones2018} making it the closest known ELM WD. Its low mass and close proximity mean that it is also one of the brightest (in an apparent sense) known ELM WDs. We demonstrate that \elm is part of a short-period, post-interacting double degenerate system with an unseen WD companion.

\section{Observations}
In this section we describe the original discovery observations and follow-up spectroscopic observations (Section~\ref{sec:spec_obs}) and survey photometric data (Section~\ref{sec:photom_obs}).

\subsection{Spectroscopic observations}
\label{sec:spec_obs}
\elm was serendipitously observed by the GALAH survey \citep{DeSilva2015,Martell:2017jd,Buder2018} on 2017 January 5 using the 3.9-metre Anglo-Australian Telescope with the HERMES spectrograph \citep{Sheinis:2015gk} and the Two-Degree Field fibre positioner top-end \citep{Lewis:2002eg}. The GALAH survey is a large spectroscopic investigation of the local stellar environment and its simple, magnitude-limited selection function includes stars such as \elm that fall outside of its normal aim of targeting unevolved disk stars. HERMES provides $\sim$1000~\AA\ of non-contiguous wavelength coverage at a spectral resolving power of $R\approx28\,000$. The raw spectra for the field containing \elm were reduced using the \textsc{2dfdr} pipeline \citep[v6.46][]{AAOSoftwareTeam2015} using the standard HERMES configuration.

As a follow-up, \elm was observed on the nights of 2019 November 5, 2019 December 16--20, and 2020 January 24--29 with the WiFeS integral field spectrograph \citep{Dopita2010} on the ANU 2.3-m telescope at Siding Spring Observatory. The instrumental setup employed the R3000/B3000 red/blue grating combination in 2019 November and the R7000/B3000 combination in subsequent runs at a nominal resolution specified by the grating label. The setup delivered a wavelength coverage of 3500--5900\AA\ in the blue and 5300--7000\AA\ (R7000) or 5300--9600\AA\ (R3000) in the red. The spectra were extracted, sky-subtracted, and wavelength calibrated using arc lamp exposures taken after stellar exposures to compensate for small wavelength shifts. The spectra were reduced using FIGARO \citep{Shortridge1993}. The reduced spectra were then (relative) flux-calibrated using observations of a number of known flux standards obtained each night.
A total of 36 individual exposures were acquired with WiFeS.

\subsection{Photometric observations}
\label{sec:photom_obs}
The \textit{Transiting Exoplanet Survey Satellite} \citep[\tess;][]{Ricker2014} is obtaining 27~d time series photometry in a sequence of sectors that will cover over 85 per cent of the sky. \elm \citep[TIC ID 43529091:][]{Stassun2019} is located in Sector 5 (observed 2018 November 15 to December 11) of the \textit{TESS} southern sky monitoring mission. It was not a target of interest, so we have only the 30~min cadence Full Frame Images. We used the \textsc{eleanor} software package \citep[][]{Feinstein2019} to download and background-subtract a $13\times13$ pixel Target Pixel File centred on \elm. We required high quality observations ($\texttt{quality}==0$), and only used cadences with dates $\mathrm{BJD}-2458438=0.78$--12.20 and 13.60--25.64 to eliminate Earthshine contamination. This provided 1080 valid observations and 52 rejections.

Table~\ref{tbl_phot} lists photometric measurements from \textit{GALEX} \citep{mor2007}, SkyMapper DR2 \citep{onk2019},
2MASS \citep{skr2006} and \textit{WISE} \citep{cut2012}.
The $NUV$ magnitude was corrected ($\Delta NUV=-0.073$) for detector non-linearity \citep{mor2007}.
\begin{table}
    \caption{Identification and measurements of \elm.}
    \centering
    \begin{tabular}{lcclcc}
    \hline
    \multicolumn{2}{l}{Parameter} & \multicolumn{3}{c}{Measurement} & Ref. \\
    \hline
    \multicolumn{2}{l}{RA (2000)} & \multicolumn{3}{c}{05$^{\rm h}$00$^{\rm m}$51\fs85} & 1 \\
    \multicolumn{2}{l}{Dec (2000)} & \multicolumn{3}{c}{-09\degr 30\arcmin 54\farcs9} & 1 \\
    \multicolumn{2}{l}{GALAH ID} & \multicolumn{3}{c}{170105002601114} & 2 \\
    \multicolumn{2}{l}{\gaia DR2 ID} & \multicolumn{3}{c}{3183166667278838656} & 3 \\
    \multicolumn{2}{l}{Parallax} & \multicolumn{3}{c}{$13.975\pm0.052$ mas} & 3 \\
    \multicolumn{2}{l}{$\mu_\alpha \cos{\delta}$} & \multicolumn{3}{c}{$-48.551\pm0.074$~mas yr$^{-1}$} & 3 \\
    \multicolumn{2}{l}{$\mu_\delta$} & \multicolumn{3}{c}{$-121.871\pm0.073$~mas yr$^{-1}$} & 3 \\
    \multicolumn{2}{l}{Period (Light curve)} & \multicolumn{3}{c}{$9.458\pm0.011$~hr} & 4 \\
    \multicolumn{2}{l}{Period (RV curve)} & \multicolumn{3}{c}{$9.462506\pm0.000017$~hr} & 4 \\
    \hline
    \multicolumn{6}{c}{Photometry} \\
    \hline
    Band & Measurement & Ref. & Band & Measurement & Ref. \\
    \hline
    $FUV$ & $14.287\pm0.009$ & 5 & $u$ & $13.347\pm0.007$ & 8 \\
    $NUV$ & $14.122\pm0.008$ & 5 & $v$ & $13.016\pm0.005$ & 8 \\
    $J$ & $12.649\pm0.026$   & 6 & $g$ & $12.514\pm0.007$ & 8 \\
    $H$ & $12.707\pm0.030$   & 6 & $r$ & $12.699\pm0.007$ & 8 \\
    $K$ & $12.709\pm0.030$   & 6 & $i$ & $12.982\pm0.005$ & 8 \\
    $W1$ & $12.628\pm0.024$  & 7 & $z$ & $13.202\pm0.005$ & 8 \\
    $W2$ & $12.689\pm0.027$  & 7 & & & \\
    \hline
    \end{tabular}\\
    References: (1) \citet{skr2006}; (2) \citet{Buder2018}; (3) \citet{Gaia2018}; (4) This work; (5) \citet{mor2007}; (6) \citet{skr2006};
    (7) \citet{cut2012}; (8) \citet{onk2019}.
    \label{tbl_phot}
\end{table}

\section{Analysis}

We proceed with a determination of the stellar parameters (Section \ref{sec:stellar_params}) of the primary component of \elm and the parameters of the orbit (Section \ref{sec:orbital_params}). Additional insights on this system are gained through an analysis of the \tess light curve (Section~\ref{sec:tess_lc}). 

The spectroscopic time series enables an analysis of the orbital parameters as well as the stellar parameters of the primary star. We found no spectroscopic signatures that would belong to the secondary star which we refer to as the unseen companion.

\subsection{Atmospheric and stellar parameters}
\label{sec:stellar_params}
We derived the physical properties of \elm using a combination of spectroscopic, photometric and astrometric data. We employed a grid of hydrogen-rich model atmospheres allowing for traces of heavier elements (He, C, N, O, Si, Ca, Fe). The models allow for convective and radiative energy transport. Model convergence is achieved with conservation of the total flux ($F_{\rm convective}+F_{\rm radiative}$) within 0.1 per cent error. The same models were used in the analysis of the ELM NLTT~11748 \citep{kaw2009}
but with updated Balmer line profiles as described in \citet{kaw2012}. 

\begin{figure}
    \centering
    \includegraphics[viewport=100 150 575 710,clip,width=\columnwidth]{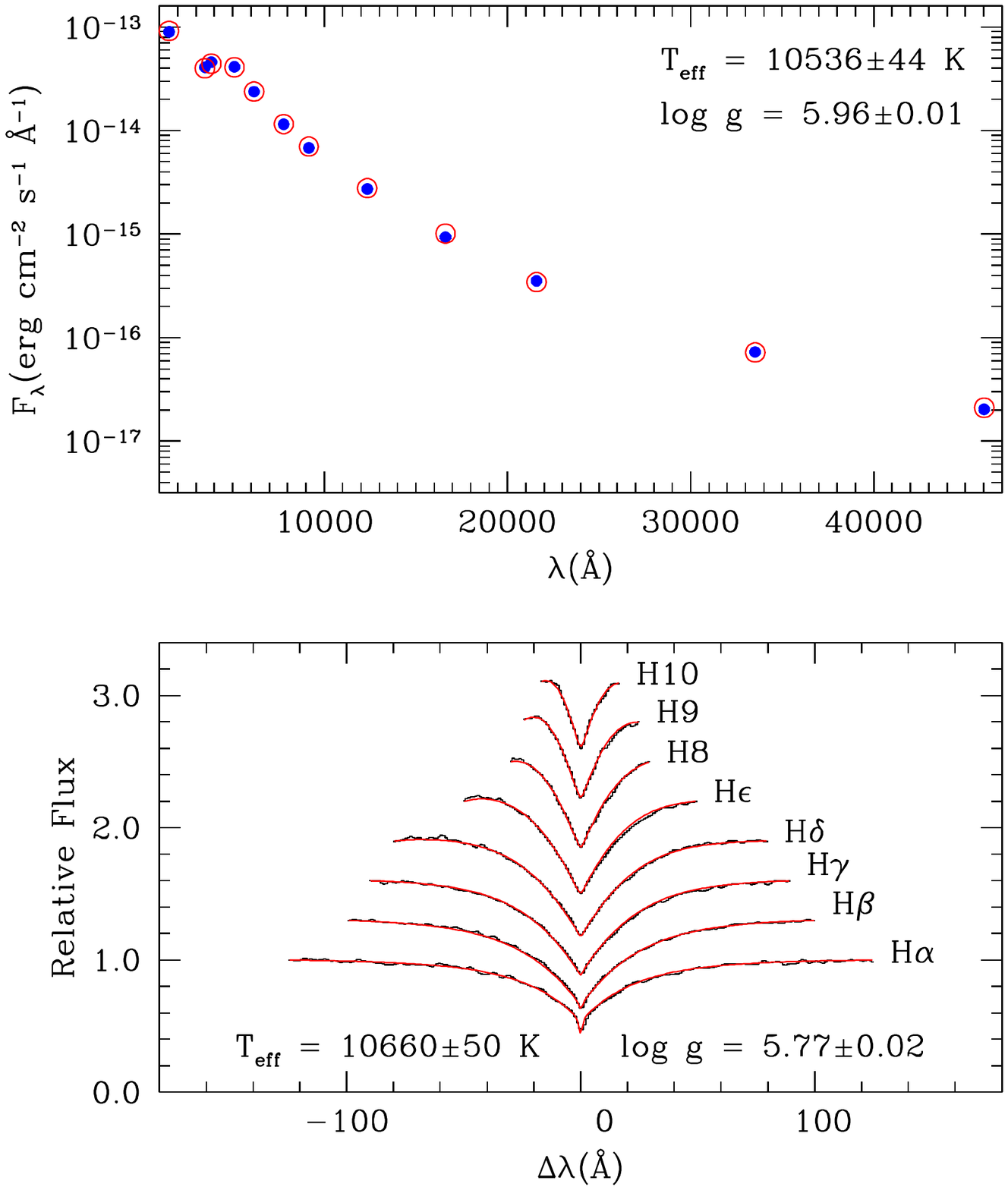}
    \caption{Top panel: The observed spectral energy distribution (blue points) compared to the best fitting
    magnitudes (open red circles). Bottom panel: Representative Balmer line spectrum (black) compared to the corresponding best fitting
    model spectrum (red).}
    \label{fig_spec_sed}
\end{figure}

First, we fitted the Balmer line profiles from H$\alpha$ to H$_{10}$ varying \teff\ and \logg. The line profiles are shaped by the temperature and density structure of the atmosphere and are dominated by Stark broadening. The upper Balmer lines are also affected by pressure ionization effects which are modelled using an energy level dilution scheme \citep{hum1988} with details provided by \citet{kaw2006}. Fig~\ref{fig_spec_sed} shows a representative result of the line profile fitting procedure with the corresponding best fit parameters. A set of 36 (\teff, \logg) measurements was obtained from the WiFeS spectra from which we computed the weighed average and dispersion:
\begin{displaymath}
\teff=10\,650\pm110\ {\rm K},\ \logg=5.76\pm0.05\ \ \ \ \ \ {\rm (Balmer)}.
\end{displaymath}
The measurements do not correlate with the orbital phase with a Pearson correlation coefficient smaller than $\approx 0.2$. 
Next, we constrained \logg\ using the \gaia DR2 distance measurement, $d_{\rm Gaia}$. The surface gravity is measured indirectly using mass-radius relations \citep{Serenelli2001,Istrate2016} and a constraint on the stellar radius $R$ set by the observed absolute magnitude $M_{\rm obs}=m-(5\log d_{\rm Gaia}-5)$ for a given photometric band. Synthetic absolute magnitudes, $M_{\rm mod}\,(\teff,\logg$), are obtained by folding the model spectra with appropriate bandpasses. Equating $M_{\rm mod}=M_{\rm obs}$, we solve numerically for \logg\ at a given \teff. Therefore, the observed Balmer spectra are fitted as a function of \teff\ with \logg\ constrained by the \gaia parallax. In the present analysis with adopted the SkyMapper $g$ band. The weighed average and dispersion of the set of 36 (\teff, \logg) measurements are:
\begin{displaymath}
\teff=10\,390\pm60\ {\rm K}, \ \logg=5.93\pm0.01\ \ \ \ {\rm (Balmer-\gaia)}.
\end{displaymath}
Finally, we fitted the spectral energy distribution (SED) mapped by a set of photometric measurements, $m_i$ (Table~\ref{tbl_phot}). Again, the apparent magnitudes are converted  into absolute magnitudes, $M_{{\rm obs},i}$, and fitted to the set of synthetic absolute magnitudes (computed as above), $M_{{mod},i}$, as a function of \teff. The best fitting parameters are:
\begin{displaymath}
\teff=10\,536\pm44\ {\rm K}, \ \logg=5.96\pm0.01\ \ \ \ {\rm (SED-\gaia)}.
\end{displaymath}
The SED does not show evidence of the secondary star, which is most likely a fainter WD.
The solutions constrained by the \gaia distance measurement are marginally consistent, but show systematic differences with \logg\ measurements obtained fitting the Balmer line profiles alone. All solutions are model dependent: the \gaia parallax measurement constrains the stellar radius, and, indirectly \logg\ by applying mass/radius relations; model Balmer line profiles that constrain \logg\ depend on pressure broadening and energy level prescriptions.

\begin{figure}
    \centering
    \includegraphics[viewport=20 160 580 690,clip,width=\columnwidth]{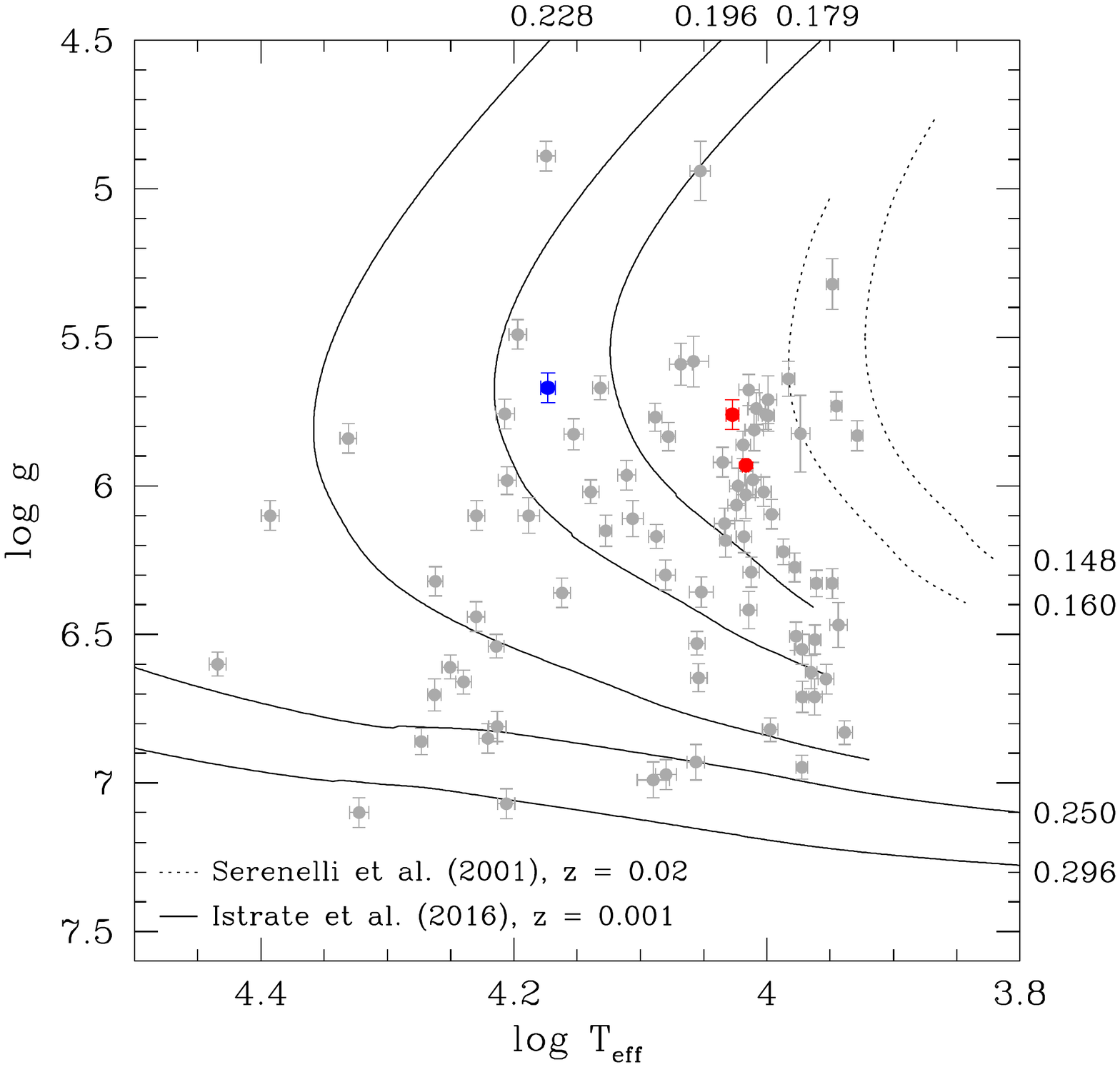}
    \caption{Effective temperature and surface gravity measurements of known ELM WDs are shown with 
    grey symbols. \elm (Balmer line profile fits with and without \gaia) and GALEX~J1717+6757 are shown with red and blue circles, respectively. The ELM WDs are compared to cooling tracks (see text) labelled with the mass (\msun).}
    \label{fig_teff_grav}
\end{figure}

A close examination of the blue spectrum of \elm\ does not show evidence of the \ion{Ca}{ii} H\&K doublet. Adopting stellar parameters obtained with the joint SED-\gaia data set and varying the calcium abundance we measured an abundance (by number) upper limit (99 per cent) of $\log\,{\rm Ca/H} \lesssim -10.7$. The calcium abundance upper-limit indicates that \elm\ does not belong to the group of high-metallicity ELM WDs described by \citet{Gianninas2014}.
Using evolutionary tracks for ELM WDs from \citet{Serenelli2001} and \citet{Istrate2016} and combining the (\teff,\logg) measurements described above we calculate a mass of $0.17\pm0.01$~\msun. Fig.~\ref{fig_teff_grav}
compares 2MASS~J0500-0930 to other known ELM WDs \citep{Brown2020,Vennes2011} and to the cooling tracks of \citet{Serenelli2001} and \citet{Istrate2016}. The stellar parameters place \elm\ amongst the lowest-mass WDs.

\subsection{Period analysis}
\label{sec:orbital_params}
\begin{figure}
	\includegraphics[viewport=15 145 570 710,clip,width=\columnwidth]{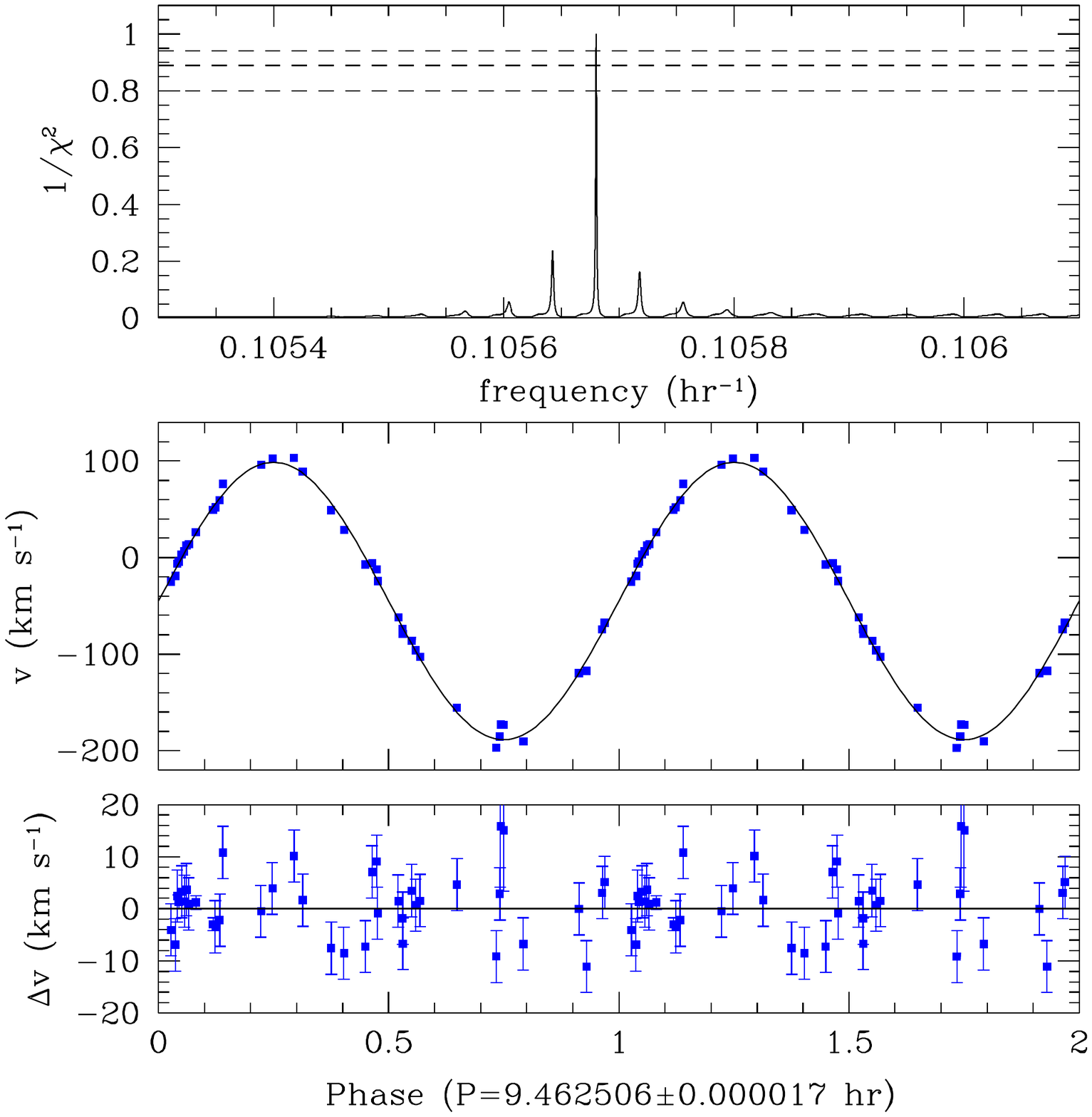}
	\caption{Period analysis of radial velocity measurements. The top panel shows
	$\chi^2$ function as a function of the frequency in cycles per hour. The horizontal dashed lines
	represent confidence levels at 66 (top), 90 and 99 per cent (bottom). The middle panel shows the
	radial velocity measurements (squares) folded on the orbital period and best fitting sine curve, and the
	bottom panel shows the radial velocity residuals.}
    \label{fig:wifes_rv}
\end{figure}

We measured the radial velocity of the ELM WD in HERMES and WiFeS spectra by fitting a Gaussian function to  the deep, narrow Doppler core ($\pm2$\AA) of H$\alpha$. The mid-exposure HJD times and corresponding radial velocities in the heliocentric rest frame are provided in the Supplementary Material. We fitted the data with a sinusoidal function
\begin{displaymath}
v(t) = \gamma + K\,sin[2\pi(t-T_0)/P]
\end{displaymath}
where $P$ is the orbital period, $T_0$ is the epoch of inferior conjunction for the ELM, $\gamma$ is the systemic velocity and $K$ is the
ELM semi-amplitude velocity. Individual error bars were set at approximately one tenth of a resolution element corresponding to the rms of the fitting procedure, which dominates the error budget, and providing increased weight to higher resolution spectra.
Fig.~\ref{fig:wifes_rv} shows the results of the analysis that delivered $P$ and $T_0$:
\begin{displaymath}
P = 0.3942711\pm0.0000007\ {\rm d}\ (= 9.462506\pm0.000017\ {\rm h}),
\end{displaymath}
\begin{displaymath}
T_0\, ({\rm HJD}) = 2458318.0472\pm0.0004.
\end{displaymath}
The ELM motion is described by $K=143.7\pm0.9~\kms$ and $\gamma=-45.0\pm1.2~\kms$. Velocity residuals averaged 5~\kms which validates the velocity error bars.

\subsection{\tess Light curve}
\label{sec:tess_lc}
\begin{figure}
	\includegraphics[width=\columnwidth]{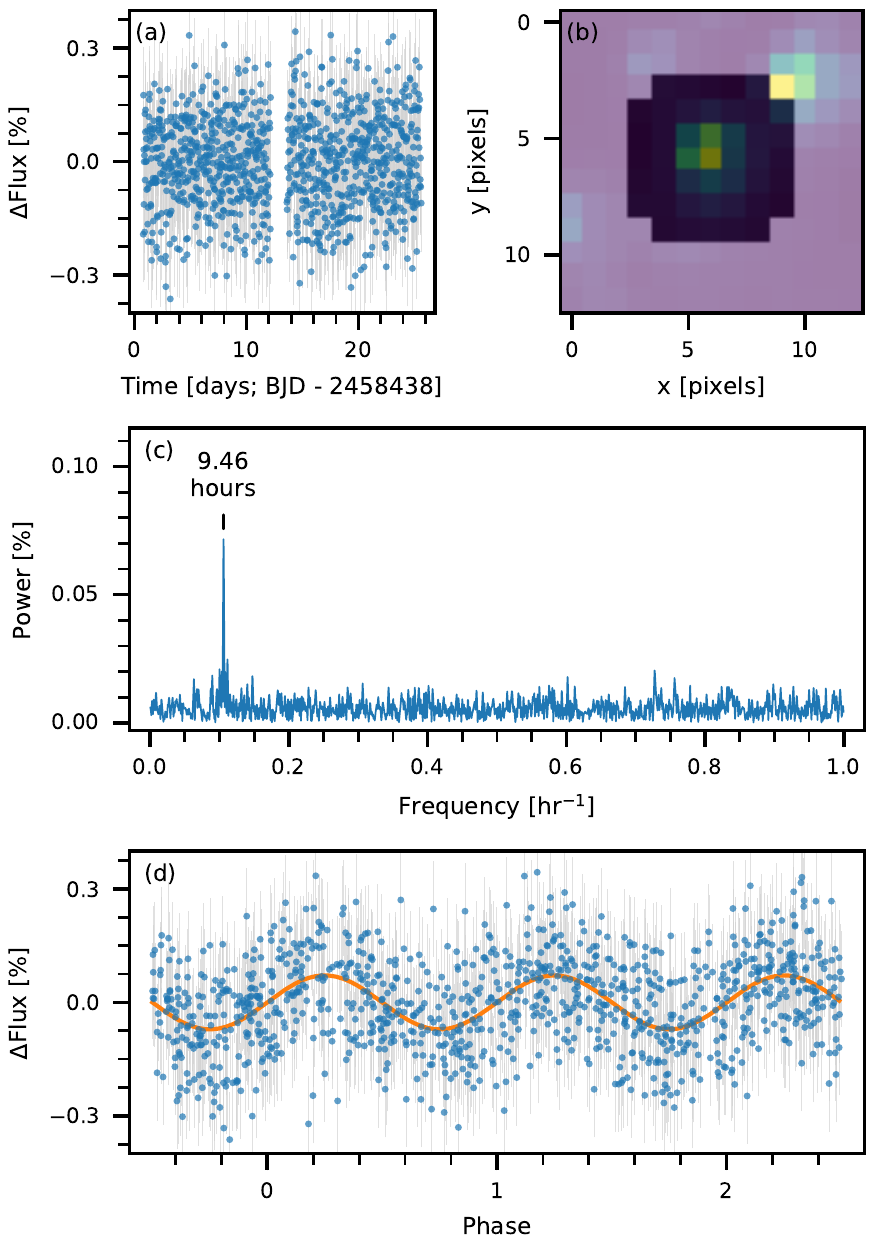}
    \caption{(a) the light curve measured by \textsc{eleanor}. (b) a 13-by-13 pixel cutout of a TESS full frame image, with the aperture used by \textsc{eleanor} indicated as the darker region. (c) the periodogram of the light curve, with the period of the peak frequency indicated (9.46~h) (d) the light curve folded on a period of 9.458~h, showing a sinusoidal behaviour. The orange curve is a sine function with a period of 9.458~h and amplitude of 0.070~per~cent.}
    \label{fig:tess_elm_light_curve}
\end{figure}
ELM WDs have been found to show evidence for eclipses, Doppler beaming, ellipsoidal, and reflection effects in their light curves \citep{Faigler2011}. A light curve from the \tess photometry was constructed with \textsc{eleanor} with a PCA-based detrending of the raw aperture photometry and modelling the PSF of the star at each cadence to generate a light curve. A normalized light curve was found with flux errors of $\approx$0.22~per cent  (Fig.~\ref{fig:tess_elm_light_curve}). The periodogram was built using \textsc{lightkurve} \citep[v1.6.0;][]{LightkurveCollaboration2018,LightkurveCollaboration2019} and the default Lomb-Scargle method \citep{Lomb1976,Scargle1982,Press1989} implemented using \textsc{astropy} \citep{TheAstropyCollaboration2018}. This showed one clear peak at $\sim$9.46~h with an amplitude of $\sim$0.071~per~cent --- very similar to the orbital period found from the RV observations.

To quantify the uncertainties of these values, we used \textsc{emcee} to fit a sinusoidal model to the light curve. This found an amplitude of $0.070\pm0.010$~per~cent, a period of $9.458\pm0.011$~h, and a $T_0\,{\rm (BJD)} = 2458438.114\pm0.017$.
The light curve does not show any evidence for eclipses deeper than 0.03 per cent at the observed 30 min cadence. At an inclination of 90$\degr$, an eclipse depth of 5 per cent with a duration of $\approx3$~min, or 0.5 per cent at a 30 min cadence, would have been observed assuming radii of 0.077 and 0.017~\rsun\ for the ELM and a cooler, 0.3~\msun\ WD companion, respectively. Shorter, shallower eclipses remain detectable at inclinations $\gtrsim87$\degr.

The period of the light curve is in agreement with the orbital period. When phasing the
light curve on the orbital period it shows evidence for the Doppler beaming effect which is caused by light concentration in the direction of motion. We observed the flux maximum at an epoch corresponding to orbital phase $\Phi_{\rm orb} = 0.78\pm0.04$ in agreement with the predicted $\Phi_{\rm orb} = 0.75$ (Fig.~\ref{fig:wifes_rv}). Since the ELM WD outshines its higher mass WD companion, the amplitude of the beaming effect can be estimated using $A \approx \alpha e^\alpha /(e^\alpha -1) \times (K/c)$ \citep{vanKerkwijk2010}, where $c$ is the speed of light, $K$ is the velocity semi-amplitude, and $\alpha = h \nu/kT_{\rm eff}$, where $\nu$ is the frequency of the photometric bandpass. In the \tess bandpass we expect an $\approx$1~mmag amplitude for the beaming effect compared to an observed amplitude of $0.77$~mmag. The concurrence of the predicted and observed phases and amplitudes affirms the Doppler beaming effect as the explanation for the variability.

\section{Discussion and Summary}

With a mass function of 0.12~\msun, the minimum companion mass at an inclination $i=90^\circ$ is 0.3~\msun, assuming an ELM mass of 0.17~\msun (see Section~\ref{sec:stellar_params}). Conversely, if the companion is a normal 0.6~\msun WD, the inclination would be $\approx$45\degr.

With the systemic radial velocity of the system calculated (see Section~\ref{sec:orbital_params}), we now have the 6D space information, from which we calculate its Galactic orbit properties. As in \citet{Simpson2017c} the orbit was calculated using \textsc{galpy} with the default options and with the Milky Way potential defined by \texttt{MWPotential2014}. We find that \elm is on a disk-like orbit, with peri-orbit of $r_\mathrm{perio}=6.73\pm0.01$~kpc, apo-orbit $r_\mathrm{apo}=11.61\pm0.09$~kpc, an eccentricity of $e=0.27\pm0.01$, and a maximum distance above the plane of  $z_\mathrm{max}=0.02\pm0.01$~kpc. About two-thirds of ELM are found in the disk \citep{Brown2020}.

Future gravitational wave detectors will have the ability to detect some double-generate systems with $P<1$~hr. Although \elm is not in this regime, we decided to calculate the characteristic strain, as this value is dependent on the inverse of the distance. Using equation 2 from \citet{Brown2020}, and assuming a fixed mass of the ELM of 0.17~\msun, for the reasonable range of secondary masses ($0.3<M_2<0.6$~\msun), the gravitational wave strain ($h_c$) for a four-year \textit{LISA} mission will be in the range of $0.5\times10^{-20}<h_c<2.0\times10^{-20}$. This unfortunately is several orders of magnitude below the detection limit.

It will take a few tens of Gyrs for 2MASS~J0500-0930 to merge through the loss of gravitational radiation \citep{Ritter1986}. More 
specifically the ELM WD would merge with a companion of a mass of 0.6~\msun\ in 36 Gyr. \citet{Brown2020} show that most ELM WD 
binaries will merge through unstable mass transfer \citep{Shen2015} rather than become AM CVn binaries.

The GALAH survey will cover up to approximately half of the sky at a limiting magnitude of $V_{\rm lim}=14$ \citep{DeSilva2015}, hence it probed so far a volume $\lesssim 0.5\,(4\pi/3)\,d_{\rm max}^3$, where $\log\,d_{\rm max}=(V_{\rm lim}-M_V+5)/5$. With an absolute magnitude $M_V\approx 8.2$, \elm\ adds a dominant contribution ($\gtrsim 160$\,kpc$^{-3}$) to the local space density of ELM WDs which is comparable to the density estimated by \citet{Brown2016b}.

In summary, we have confirmed \elm as the closest ELM WD to the Sun at a distance of 71~pc, while the next one was found at 178~pc \citep[GALEX~J1717+6757, ][]{Vennes2011}. The minimum companion mass of 0.3~\msun\ and the lack of infrared excess indicate that the companion is a fainter WD with a higher mass and lower effective temperature. The \tess light curve does not show evidence of an eclipse and is modulated over the orbital period by the Doppler beaming effect. Its kinematics show that \elm belongs to the disk population. Finally, \elm will merge with its companion in a time $>t_{\rm Hubble}$ with the likelihood of a Type Ia supernova event dependent on the secondary mass.

\section*{Acknowledgements}
The GALAH survey is based on observations made at the Anglo-Australian  Telescope, under programmes A/2013B/13, A/2014A/25, A/2015A/19, A/2017A/18. We acknowledge the traditional owners of the land on which the AAT stands, the Gamilaraay people, and pay our respects to elders past, present and emerging. This paper includes data that have been provided by AAO Data Central (\url{datacentral.org.au}).

This paper includes data collected by the TESS mission funded by the NASA Explorer Program and data from the European Space Agency (ESA) mission {\it Gaia} (\url{https://www.cosmos.esa.int/gaia}), processed by the {\it Gaia} Data Processing and Analysis Consortium (DPAC, \url{https://www.cosmos.esa.int/web/gaia/dpac/consortium}).

SV thanks the International Centre for Radio Astronomy Research for their support.
AK and SV thank the Australian National University for their support.
JDS acknowledges the support of the Australian Research Council (ARC) through Discovery Project grant DP180101791. MSB and GSDC acknowledge ARC support through Discovery Project grant DP150103294. A.F.M. received funding from the European Union's Horizon
2020 research and innovation programme under the Marie Sklodowska-Curie grant agreement 797100.

The following software and programming languages made this research possible: \textsc{python} (v3.7.6); \textsc{astropy} \citep[v4.0;][]{TheAstropyCollaboration2018}, a community-developed core Python package for astronomy; \textsc{matplotlib} \citep[v3.1.3;][]{Hunter2007,Caswell2020}; \textsc{scipy} \citep[v1.4.1;][]{SciPy1.0Contributors2020};  \textsc{galpy} \citep[v1.5; \url{http://github.com/jobovy/galpy};][]{Bovy2015}; \textsc{eleanor} \citep[v1.0.1;][]{Feinstein2019}; \textsc{emcee} \citep[v3.0.2;][]{Foreman-Mackey2013,Foreman-Mackey2013a}.

%%%%%%%%%%%%%%%%%%%%%%%%%%%%%%%%%%%%%%%%%%%%%%%%%%



\bibliographystyle{mnras}
\bibliography{library} % if your bibtex file is called example.bib

%%%%%%%%%%%%%%%%%%%%%%%%%%%%%%%%%%%%%%%%%%%%%%%%%%

% Don't change these lines
\bsp	% typesetting comment
\label{lastpage}
\end{document}